\documentclass[floatfix,twocolumn,superscriptaddress, showkeys,nofootinbib]{revtex4-1}

\usepackage[hidelinks]{hyperref}
\usepackage{graphicx}%
\usepackage{multirow}%
\usepackage{amsmath,amssymb,amsfonts}%
\usepackage{mathrsfs}%
\usepackage[title]{appendix}%
\usepackage{xcolor}%
\usepackage{textcomp}%
\usepackage{booktabs}%
\usepackage{algorithm}%
\usepackage{algorithmicx}%
\usepackage{algpseudocode}%
\usepackage{listings}%
\usepackage{xurl}%
\usepackage{url}
\usepackage{braket}
\usepackage[misc,geometry]{ifsym}

\usepackage[caption=false]{subfig}

\begin{document}

\title{Benchmarking Quantum Generative Learning:\\ A Study on Scalability and Noise Resilience using QUARK}

\author{Florian J. Kiwit}
\email{florian.kiwit@bmw.de}
\affiliation{BMW Group, Munich, Germany}
\affiliation{Ludwig Maximilian Universität, Munich, Germany}

\author{Maximilian A. Wolf}
\affiliation{BMW Group, Munich, Germany}
\affiliation{Ludwig Maximilian Universität, Munich, Germany}

\author{Marwa Marso}
\affiliation{BMW Group, Munich, Germany}
\affiliation{Ludwig Maximilian Universität, Munich, Germany}
\affiliation{Technical University, Munich, Germany}

\author{Philipp Ross}
\affiliation{BMW Group, Munich, Germany}

\author{Jeanette M. Lorenz}
\affiliation{Ludwig Maximilian Universität, Munich, Germany}
\affiliation{Fraunhofer Institute for Cognitive Systems IKS, Munich, Germany}

\author{Carlos A. Riofr\'io}
\affiliation{BMW Group, Munich, Germany}

\author{Andre Luckow}
\affiliation{BMW Group, Munich, Germany}
\affiliation{Ludwig Maximilian Universität, Munich, Germany}


\begin{abstract}
Quantum computing promises a disruptive impact on machine learning algorithms, taking advantage of the exponentially large Hilbert space available. However, it is not clear how to scale quantum machine learning (QML) to industrial-level applications. This paper investigates the scalability and noise resilience of quantum generative learning applications. We consider the training performance in the presence of statistical noise due to finite-shot noise statistics and quantum noise due to decoherence to analyze the scalability of QML methods. We employ rigorous benchmarking techniques to track progress and identify challenges in scaling QML algorithms, and show how characterization of QML systems can be accelerated, simplified, and made reproducible when the QUARK framework is used. We show that QGANs are not as affected by the curse of dimensionality as QCBMs and to which extent QCBMs are resilient to noise.
\end{abstract}

\maketitle

\section{Introduction}\label{sec:intro}

Systematic evaluation of quantum processors and algorithms through benchmarking offers valuable insights into the current capabilities and future potential of available quantum processing units~\cite{Proctor2022,Erhard2019,BlumeKohout2020volumetricframework,Mills2021application,10.1145/3464420,Lubinski_etal_2021,lubinski2024quantum,Finzgar_2022,bowles2024better}. However, benchmarking quantum computing is far from a straightforward task. The field is characterized by a diversity of technologies~\cite{Ladd2010}, each with unique requirements for precise and meaningful assessment. As a result, current benchmarks often focus on specific aspects of the technology, which can sometimes lead to an incomplete picture of the end-to-end performance of quantum computing.

The Quantum computing Application benchmark (QUARK) framework~\cite{Finzgar_2022} was explicitly developed for challenges of application-oriented quantum computing. QUARK's benchmarking approach ensures a comprehensive evaluation, covering the entire benchmarking pipeline from hardware to algorithmic design for the problems under investigation. Its versatility and modular implementation are central to QUARK, allowing for component expansion and customization. Additionally, it hosts benchmarks from the domain of optimization~\cite{Finzgar_2022} and machine learning~\cite{kiwit2023applicationoriented}.

In quantum machine learning (QML), scaling algorithms and maintaining performance amidst noise are crucial for practical applications, particularly in industries reliant on generative models. This work shows an extension that enables us to include noisy simulations for QML applications. It is important to understand the limitations and track the development of current quantum hardware and algorithms over time. Concretely, we present a comprehensive study of the scalability of QML models, evaluating their intrinsic robustness against statistical and quantum noise. We aim to bridge the gap between theoretical QML advancements and their practical implementation in real-world scenarios while using the QUARK framework to accelerate and standardize performance assessment.

The paper is organized as follows: Section~\ref{QML_basics} briefly introduces the basic concepts of quantum generative learning used in this paper. In Section~\ref{QUARK_The_framwork}, we present a description of the components of the QUARK framework. In Section~\ref{sec:convergence}, we discuss the influence of statistical noise on the scalability of QML models. In Section~\ref{sec:noise}, we look at the effects of quantum noise and hardware defects on the training of QML models. And finally, in Section \ref{Conclusion}, we summarize our findings.

\section{Quantum Generative Learning}\label{QML_basics}
Generative modeling is a growing area of interest across all industries. Applications include anomaly detection, text \& image generation, or speech \& video synthesis. Ultimately, the objective of training a generative model is to express the underlying distribution of a dataset by a machine learning model. In QML, this model is represented by a parameterized quantum circuit (PQC)~\cite{Benedetti_2019_2}. During the training of a quantum generative model, the probability amplitudes of the quantum state vector generated by a PQC are fitted to the probability distribution of the dataset; see, for example, reference~\cite{schuld_machine_2021}. We will refer to the probability mass function (PMF) of the dataset as $p$ and to that of the state generated by the PQC as $q$. The absolute square values of the state vector give the PMF of the PQC.

Two popular training routines are the \emph{quantum circuit Born machine} (QCBM) \cite{benedetti2019generative} and the \emph{quantum generative adversarial network} (QGAN) \cite{PhysRevLett.121.040502}, see~\cite{riofrio2023performance} for a detailed review of both methods. The QCBM tries to minimize the Kullback-Leibler (KL) divergence, a well-known statistical distance, between $p$ and $q$ by adapting the model parameters via the covariance matrix adaption evolutionary strategy (CMA-ES)~\cite{hansen2019pycma}, a gradient-free optimizer.

Conversely, the QGAN follows the architecture of a classical GAN~\cite{NIPS2014_5ca3e9b1}. A GAN operates on the principle of adversarial training in a minimax 2-player game, employing two neural networks, the generator and the discriminator. The generator creates synthetic data instances, while the discriminator evaluates these generated samples alongside real ones. The two networks engage in a continual feedback loop, with the generator striving to improve its output and the discriminator refining its ability to differentiate between real and fake samples. Different architectures are proposed for QGANs, but a typical approach is to replace the classical generator with a PQC. The model parameters of the PQC are updated by gradient descent, and the parameter-shift rule determines the gradients, for example, reference~\cite{PhysRevA.99.032331}. The classical discriminator of the QGAN is optimized with the ADAM~\cite{kingma2017adam} optimizer, and the gradients are determined via backpropagation.

\section{Benchmarking Quantum Computing}
In its most general form, benchmarking is the process of comparing the performance of systems by a set of measurements. The workloads used are referred to as benchmarks~\cite{jain1991art}, and the metrics are the criteria to compare the performance. In quantum computing, these metrics should characterize scale, quality, and speed~\cite{amico2023defining}. 
Depending on the scope of attributes a benchmark assesses, they can be attributed to three categories~\cite{wang2022sok}. (1) Physical Benchmarks focus on the basic physical properties of quantum hardware, such as the number of qubits and quantum gates. Examples are T1 and T2 relaxation times, gate fidelity, and readout fidelity; for example, see reference~\cite{Eisert2020} for an extensive review. (2) Aggregated benchmarks assess the performance over a large set of device attributes. Prominent examples are quantum volume~\cite{Cross2019} and circuit layer operations per second~\cite{wack2021quality}. (3) Application-oriented benchmarks test the performance of quantum computers in real-world scenarios. They simulate specific computational tasks that quantum computers are expected to perform, e.\,g., optimization, machine learning and chemistry algorithms. These benchmarks are particularly important for demonstrating the practical utility and efficiency of quantum systems in solving complex, real-world problems and are a key indicator of progress toward quantum advantage. Examples include QPack~\cite{koen_qaoa_benchmarking_2022}, QED-C~\cite{Lubinski_etal_2023,Lubinski_etal_2021,lubinski2024quantum} and Q-Score~\cite{Martiel_2021}.

\subsection{The QUARK Framework}
\label{QUARK_The_framwork}

\begin{figure*}[!bt]
    \centering
    \includegraphics[width=0.68\linewidth]{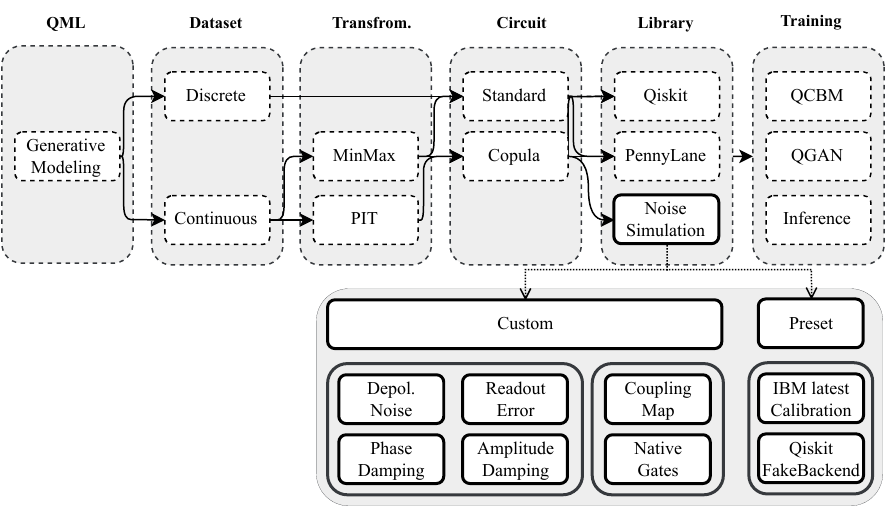}
    \caption{Illustration of the components of the QUARK framework for quantum generative modeling with a detailed depiction of the noise module. We start at the left by defining the application, in this case, the training of a QML model. Then, the user defines the dataset for training, followed by necessary data transformations, before choosing the PQC ansatz to be used as a model. As explained in the main text, QUARK offers great flexibility regarding quantum libraries for implementing the circuit ansatz and training via QCBM or QGAN methods. At the bottom, we show the structure of the new Noise Simulation module. This module is designed to offer both predefined noise configurations and the flexibility to create entirely custom noise profiles and backend specifications. This allows for more accurate assessments of the robustness and performance of quantum algorithms under various noise conditions. The new modules are indicated by bold solid lines in the figure. More details are given in the main text.}
    \label{fig:QUARK structure}
\end{figure*}

The QUARK framework~\cite{quarkGithub} orchestrates application-oriented quantum benchmarks in a well-defined, standardized, reproducible and verifiable way. It remains vendor-neutral to ensure unbiased application across different quantum computing platforms. In the following, we will describe the implementation of the generative modeling application in QUARK. 

A benchmarking instance is defined by configuring six distinct modules (see Figure~\ref{fig:QUARK structure}): In the (1) generative modeling class, global properties, such as the number of qubits $n$, are defined. Subsequently, a (2) continuous or discrete dataset is selected. The discrete datasets are characterized by a constraint on the bit string of length $n$, while the continuous datasets include both low-dimensional synthetic and real data. The continuous datasets are passed to a (3) transformation. This ensures the data is in a standard and normalized form. The MinMax transformation maps the marginal distributions to the interval $[0, 1]$. An alternative is the probability integral transformation (PIT), which makes the marginal distributions uniformly distributed. After applying the transformation, the data is mapped to a discrete probability distribution with $2^n$ bins. Next, the architecture of the PQC is selected in the (4) circuit module and mapped to the (5) Qiskit~\cite{Qiskit} or PennyLane~\cite{bergholm2018pennylane} SDK. The library-agnostic definition of the PQC enables comparative studies of quantum simulators from different vendors. Additionally, many variational ansatzes can be extended as the application requires. For example, the \textit{copula circuit}~\cite{PhysRevResearch.4.043092}, which naturally respects the properties of data transformed via the PIT, can only learn a probability distribution whose cumulative marginals are uniformly distributed. In the last step, a pre-trained PQC is loaded for (6) inference or a training routine is configured. Training routines include the QCBM and QGAN, discussed in the previous section, but can be extended to other QML models. After defining the benchmarking instance, QUARK orchestrates the execution, data collection and visualization of the benchmark. For a detailed report on the QUARK framework, see References~\cite{Finzgar_2022,kiwit2023applicationoriented}. 

Furthermore, both noisy and noise-free simulators are available. Depending on the individual requirements, one can configure the backend of the circuit to include multiple sources of errors. This can be done in two ways: (1) By specifying different error sources, like readout and depolarizing errors or amplitude and phase damping and chip-agnostic parameters, e.\,g., the coupling map, which represents the connectivity of the qubits; the basis gates, i.e., the gates that can be used on the backend. (2) We also provide an implementation of Qiskit FakeBackends, which are a predefined snapshot of the error rates, coupling map and basis gates ~\cite{QiskitBackends2023}. Additionally, the latest calibration data of IBM's quantum processors can be accessed using qiskit\_ibm\_runtime. This feature enables users to understand how quantum noise might influence their selected applications and compare them to an ideal environment or a real QPU.

\section{Methodology and Results}
In this Section, we use the QUARK framework to study the effects of statistical and quantum noise in QML applications. First, we compare the training routines of the QCBM and the QGAN. Next, we study the effects of quantum noise in the training of QCBMs.

\subsection{Scalability of Quantum Generative Models under the Influence of Statistical Noise}
\label{sec:convergence}
 
The training of the QCBM relies on estimating the full PMF $q$ generated by the PQC. As we scale the number of qubits, $n$, the number of bins of the PMF scales exponentially. In the following, we will describe how the number of circuit executions, $n_\text{shots}$, scales, to keep the statistical error of the PMF, $|\alpha_i|^2$, bounded, where $\alpha_i$ are the probability amplitudes of the quantum state vector generated by a PQC. The probability of frequency $n_i$ in bin $i$ is given by the binomial distribution $B_{n_\text{shots}, p_i}(n_i)$, with $\sum n_i = n_\text{shots} $ and $p_i = n_i / n_\text{shots}$. Furthermore, $p_i$ is the estimator of $|\alpha_i|^2$. On average, if $|\alpha_i|^2$ decays exponentially with the number of qubits $n$, i.e. $|\alpha_i|^2 \propto 1/n^2 $, then the variance is proportional to $\sigma^2(p_i) \propto 2^n / n_\text{shots}$. Therefore, the number of circuit executions needs to scale exponentially with the number of qubits to keep the statistical error on each estimator of $|\alpha_i^2|$ bounded. For the QCBM, the number of circuit executions of one epoch is given by $n_\text{shots} \cdot \lambda $. The population size $\lambda$ is a hyperparameter of the CMA-ES optimizer and refers to the number of model parameters evaluated at any given iteration.

Unlike the QCBM, the number of circuit executions of the QGAN needed to update the model parameters does not increase exponentially with the number of qubits. The total number of circuit executions per epoch is given by $(2 \cdot n_\text{parameters} + 1) \cdot n_\text{samples}$. One circuit execution with the model parameters is needed to generate synthetic samples to update the discriminator. In the backward pass, two additional circuit executions per model parameter are necessary to determine the gradients with the parameter-shift rule.

\begin{figure}[tbh]
    \centering
    \includegraphics[width=\linewidth]{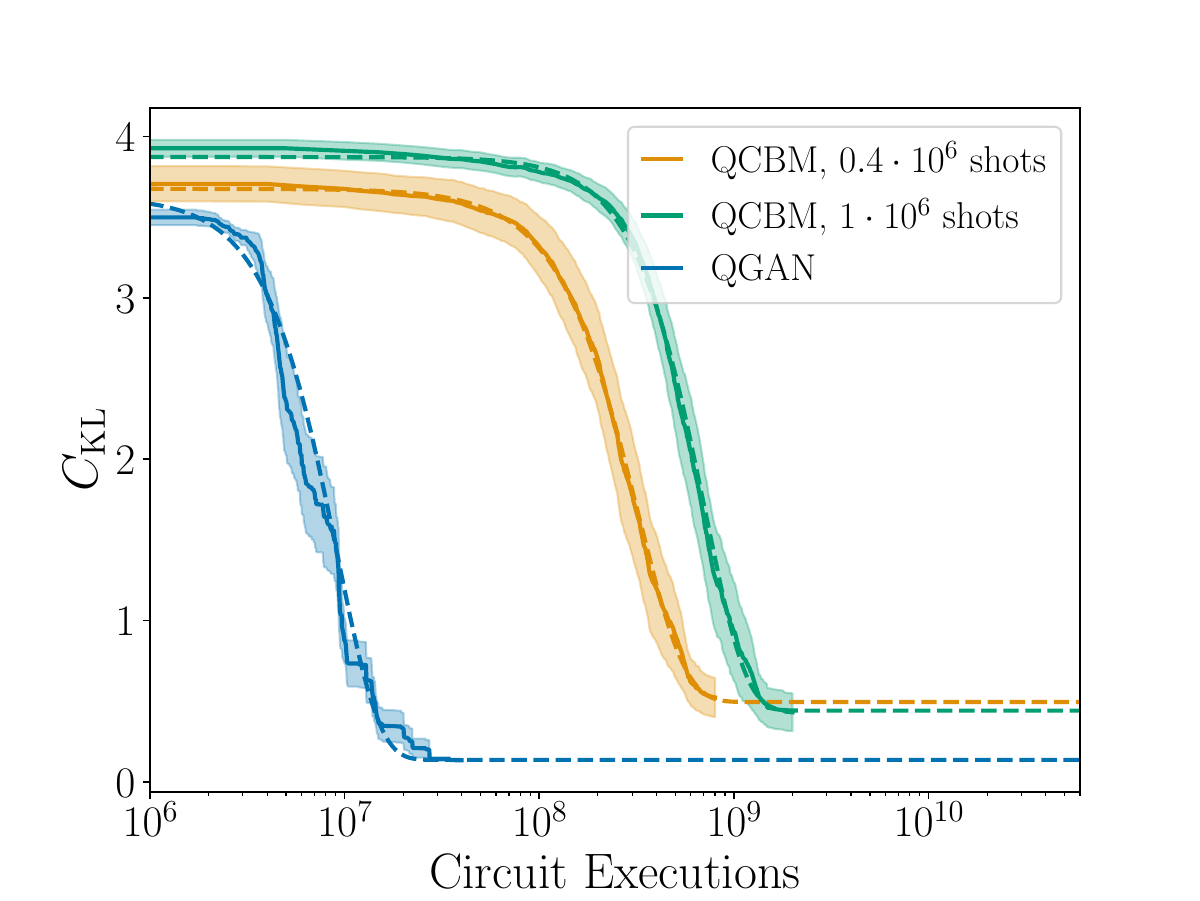}
    \caption{KL divergence, $C_\text{KL}$, as a function of the cumulative number of circuit executions for the QCBM and the QGAN. The dashed lines indicate the stretched exponential function $f(x) = \alpha \cdot \text{exp}(-\beta x^\gamma) + C_\text{KL}^\text{conv}$ fitted to the loss curves. For the training of the QCBM we show results of experiments, where we used $0.4\cdot 10^6$ and $1\cdot 10^6$ circuit executions to determine the PMF generated by the PQC. The QGAN converges with more than an order of magnitude fewer circuit executions than the QCBM to a lower limit. Each model was trained ten times and mean values and standard error on the mean $\sigma / \sqrt{10}$ are depicted as the solid lines and shaded areas, respectively.}
    \label{fig:QGAN_QCBM_Scalability}
\end{figure}

\emph{Experimental Design:} 
To showcase the different scaling behaviors of QCBMs and QGANs, we track the KL divergence as a function of the cumulative number of circuit executions, as depicted in Figure~\ref{fig:QGAN_QCBM_Scalability}. We fitted the copula circuit with 12 qubits and a depth of one to a dataset resembling the shape of the letter X, using the quantum noise-free Qiskit AerSimulator. For the QCBM, we use a population size of $\lambda=5$ and train the models with $4\cdot 10^5$ and $1\cdot 10^6$ circuit executions to determine the PMF generated by the PQC. For the QGAN, we use a batch size of 20 and alternately update the generator and discriminator on each mini-batch. While training the generative models with a shot-based simulator, we report the KL divergence between the PMF\footnote{Here we use the precise PMF, not the estimated PMF, to circumvent the influence of shot-noise when comparing the model performance.} of the PQC and the target distribution. To compare the performance of the trained models, we fit a stretched exponential function, $f(x) = \alpha \cdot \text{exp}(-\beta x^\gamma) + C_\text{KL}^\text{conv}$, to the loss curves and report the limit of the KL divergence $C_\text{KL}^\text{conv}$.

\emph{Discussion:} 
For the QCBM, increasing the number of circuit executions to estimate the PMF of the state generated by the PQC from $4\cdot 10^5$ to $1\cdot 10^6$ leads to a slight decrease of the KL divergence at convergence from $C_\text{KL}^\text{conv} = 0.49 \pm 0.13$ to $ C_\text{KL}^\text{conv} = 0.44 \pm 0.12$. The QGAN achieves faster convergence than the QCBM, requiring more than one order of magnitude fewer circuit executions. In addition to fewer circuit executions, the KL divergence of the QGAN converges to a lower limit with a value $C_\text{KL}^\text{conv} = 0.14 \pm 0.01$. To match the limit of the KL divergence of the QGAN with the QCBM, we would need to increase the circuit executions further, and the separation with respect to the circuit executions would become even more dominant.

\emph{Limitations:} The QCBM was trained with the gra\-di\-ent-free optimizer CMA-ES. However, a gradient-based training of the QCBM~\cite{liu2018} was not investigated and might lead to faster convergence despite needing more circuit evaluations per iteration to estimate the gradients. Furthermore, we used only one dataset; exploring how the findings generalize to different datasets is an interesting path for future research.

\subsection{Noisy training of QCBMs}
\label{sec:noise}

\begin{figure*}[thb]
    \centering
     \subfloat[]{\includegraphics[width=0.49\linewidth]{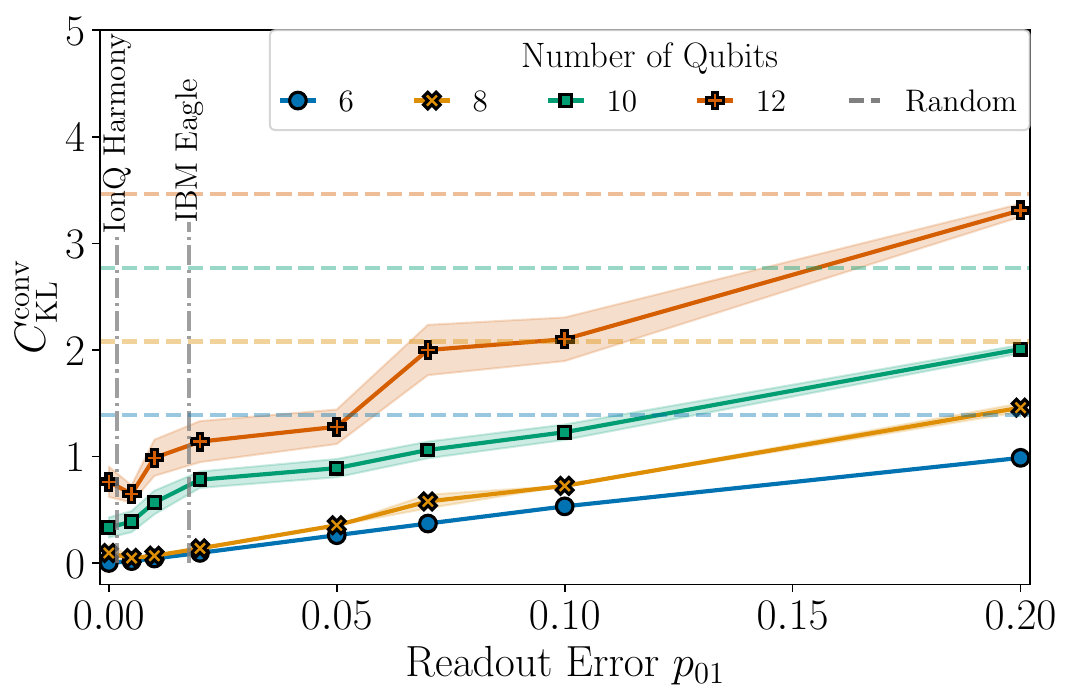}}
     \label{subfig:readout}
     \subfloat[]{\includegraphics[width=0.49\linewidth]{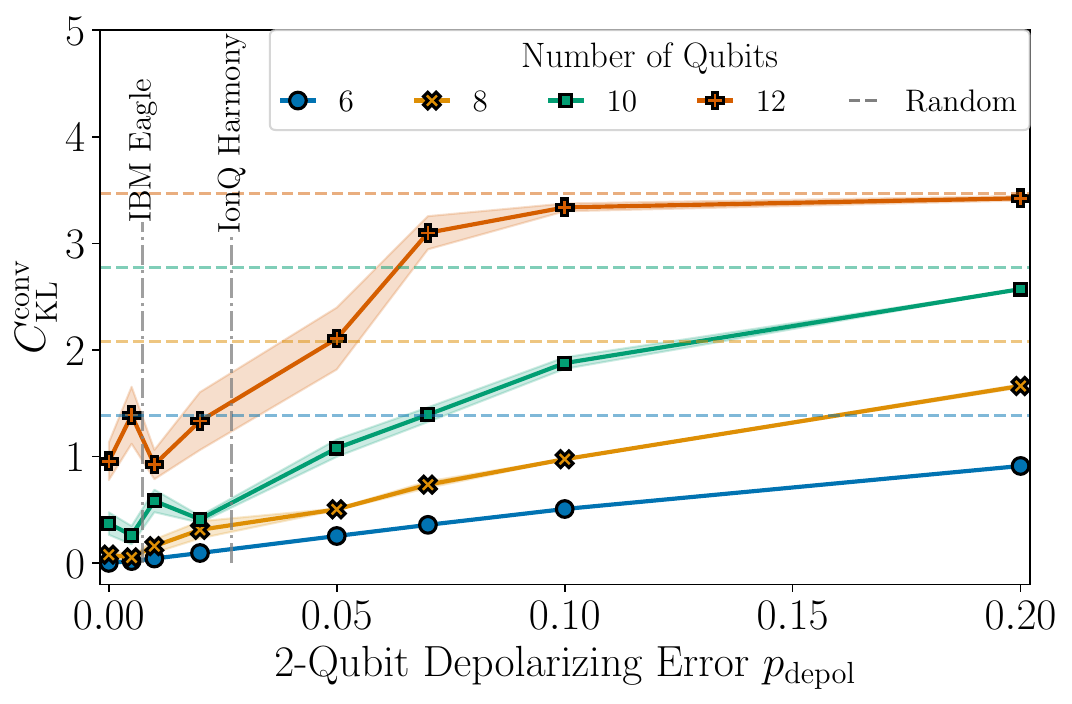}} 
    \caption{Mean KL divergence after $4\cdot10^{4}$ circuit evaluations as a function of (a) the readout error and (b) the two-qubit depolarizing error for different circuit widths of a QCBM with the standard error ($\sigma / \sqrt{8}$). The vertical grey lines denote the error value of the IBM Eagle processor (Median ECR error: $7.477\cdot 10^{-3}$ and readout error: $1\cdot10^{-2}$ of IBM Sherbrooke~\cite{IBMerrorrates}) and IonQ Harmony (SPAM error: $1.8\cdot10^{-3}$ and two-qubit gate errors: $2.7 \cdot 10^{-2}$~\cite{IonQerrorrates}). The colored horizontal lines denote the KL divergence of the training data and a uniform distribution.}
    \label{fig:quantum_noise}
\end{figure*}

In the noisy intermediate-scale quantum (NISQ) era, the performance of QML algorithms is impacted by the presence of quantum noise~\cite{preskill2018,nielsen_chuang_2010}. This noise stems from various sources, such as decoherence, imperfect gate operations, and environmental interference, and leads to a decreased fidelity of the quantum state generated by a PQC; for a detailed discussion of how noise can influence the training of PQCs, see~\cite{oliv2022evaluating}.

\begin{table}[b]
\centering
\begin{tabular}{|l|l|l|}
\hline
 $n_\mathrm{qubits}$  & \textbf{1-Qubit Gates} & \textbf{2-Qubit Gates}  \\ \hline
6  & 21              & 9                       \\ \hline
8  & 28              & 16                    \\ \hline
10 & 35             & 25                          \\ \hline
12 & 42               & 36                               \\ \hline
\end{tabular}
\caption{The total number of the one-qubit gates ($RZ$, $SX$, $RX$, $H$) and two-qubit gates ($RXX$, $CX$) of the compiled circuit for 6, 8, 10 and 12 qubits.}
\label{table:gate_count}
\end{table}

\emph{Experimental design:} We investigate the robustness of QCBMs against quantum noise to characterize the limits of current quantum hardware. To this aim, we train the QCBM and investigate the KL divergence at convergence under varying noise conditions. We fit the copula circuit with 6, 8, 10, and 12 qubits to a dataset that resembles the shape of the letter X. We execute the circuit $1\cdot 10^4$ times to estimate the PMF. The gate set used for the copula circuit is native to IonQ Harmony. The number of gates of the circuit is reported in Table~\ref{table:gate_count}. We vary the probability of (a) readout and (b) two-qubit depolarizing errors, as depicted in Figure~\ref{fig:quantum_noise}. Readout errors are represented by a bit-flip channel, which means that with a probability $p_{10}$ ($p_{01}$), the prepared input state $\ket{1}$ ($\ket{0}$) yields the measurement outcome $0$ (1). In our experiments, we set $p_{10}=p_{01}$~\cite{nielsen_chuang_2010}. Two-qubit depolarizing errors are characterized by the error rate, $p_\text{depol}$, that a two-qubit gate creates the fully mixed state instead of the desired output state of the operation.

\emph{Discussion:}
Figure~\ref{fig:quantum_noise}a illustrates the influence of readout error on the KL divergence at convergence $C_\mathrm{KL}^\mathrm{conv}$. With increasing error rate, $C_\mathrm{KL}^\mathrm{conv}$ increases linearly. Even at an error rate of 0.1, the QCBM maintains robustness against readout errors, as the $C_\mathrm{KL}^\mathrm{conv}$ is still below the random baseline\footnote{The random baseline denotes the KL divergence between the training set PMF and the uniform distribution.}. Unsurprisingly, the two-qubit depolarising error shows a stronger effect on $C_\mathrm{KL}^\mathrm{conv}$, as depicted in Figure~\ref{fig:quantum_noise}b for circuits with more gates, as multiple operations are performed per qubit, instead of measuring the state only once. The number of two-qubit gates increases with the circuit width (see Table~\ref{table:gate_count}), so does the influence of $p_\text{depol}$ on the performance of the QCBM. The performance is still below the random baseline for 6 qubits and $p_\text{depol} = 0.2$. For 12 qubits, however, the model performance corresponds to random guessing already for $p_\text{depol}=0.1$.

\emph{Limitations:} 
Our focus on readout and two-qubit depolarizing errors provides a foundation for understanding specific noise sources, but other factors, such as phase errors and crosstalk, are not considered. Our evaluation primarily considers the noise levels of state-of-the-art trapped ion quantum computers. However, the landscape of quantum hardware is rapidly evolving, and the generalizability of our findings to other quantum platforms with potentially different error characteristics needs further exploration. Incorporating mitigation and correction strategies for quantum errors could offer additional insights into enhancing the robustness of QCBMs in noisy quantum environments.

\section{Conclusion and Future Work}\label{Conclusion}
In this work, we use and extend the QUARK benchmarking framework and illustrate its functioning with two applications that consider different aspects of noise in QML: statistical and quantum noise. The modular structure of QUARK makes it a versatile tool for a broad spectrum of research applications in QML and quantum computing in general. 

Our experiments focused on the performance characterization of quantum generative models. A comparative analysis of QCBMs and QGANs revealed differences in their efficiencies. Remarkably, QGANs achieved faster convergence with reduced computational demands.
Despite the recent advancements in QPU architectures leading to reduced error rates, noise is still a limiting factor in the current NISQ era. To effectively mitigate the impact of noise in QML, strategies such as minimizing gate counts, employing QPUs with lower noise profiles, and designing circuits with inherently more resistant architectures must be pursued. 

Our studies on the influence of noise on the training of QCBMs are limited to readout and two-qubit depolarizing errors. Future studies may benefit from incorporating a more complete noise model, including noise sources such as amplitude damping or crosstalk between qubit pairs. Furthermore, extending our studies to different quantum generative models, such as QGANs, would be interesting. After simulating the influence of noise on the performance of quantum generative models, conducting the experiments on quantum hardware would be a natural step. One compelling aspect to explore is investigating if a model trained with a noisy simulator has learned to resist the noise. Based on our experience, the QUARK framework should be the tool of choice for future research.

\begin{acknowledgements}
PR and CAR were partly funded by the German Ministry for Education and Research (BMB+F) in the project QAI2-Q-KIS under Grant 13N15583. AL was partly funded by the Bavarian State Ministry of Economic Affairs in the project BenchQC under Grant DIK-0425/03.
\end{acknowledgements}

\clearpage

\bibliographystyle{unsrtnat}
\bibliography{bibliography}

\begin{thebibliography}{40}
\providecommand{\natexlab}[1]{#1}
\providecommand{\url}[1]{\texttt{#1}}
\expandafter\ifx\csname urlstyle\endcsname\relax
  \providecommand{\doi}[1]{doi: #1}\else
  \providecommand{\doi}{doi: \begingroup \urlstyle{rm}\Url}\fi

\bibitem[Proctor et~al.(2022)Proctor, Rudinger, Young, Nielsen, and Blume-Kohout]{Proctor2022}
Timothy Proctor, Kenneth Rudinger, Kevin Young, Erik Nielsen, and Robin Blume-Kohout.
\newblock Measuring the capabilities of quantum computers.
\newblock \emph{Nature Physics}, 18\penalty0 (1):\penalty0 75--79, 1 2022.
\newblock ISSN 1745-2481.
\newblock \doi{10.1038/s41567-021-01409-7}.
\newblock URL \url{https://doi.org/10.1038/s41567-021-01409-7}.

\bibitem[Erhard et~al.(2019)Erhard, Wallman, Postler, Meth, Stricker, Martinez, Schindler, Monz, Emerson, and Blatt]{Erhard2019}
Alexander Erhard, Joel~J. Wallman, Lukas Postler, Michael Meth, Roman Stricker, Esteban~A. Martinez, Philipp Schindler, Thomas Monz, Joseph Emerson, and Rainer Blatt.
\newblock Characterizing large-scale quantum computers via cycle benchmarking.
\newblock \emph{Nature Communications}, 10\penalty0 (1):\penalty0 5347, 11 2019.
\newblock ISSN 2041-1723.
\newblock \doi{10.1038/s41467-019-13068-7}.
\newblock URL \url{https://doi.org/10.1038/s41467-019-13068-7}.

\bibitem[Blume-Kohout and Young(2020)]{BlumeKohout2020volumetricframework}
Robin Blume-Kohout and Kevin~C. Young.
\newblock A volumetric framework for quantum computer benchmarks.
\newblock \emph{{Quantum}}, 4:\penalty0 362, November 2020.
\newblock ISSN 2521-327X.
\newblock \doi{10.22331/q-2020-11-15-362}.
\newblock URL \url{https://doi.org/10.22331/q-2020-11-15-362}.

\bibitem[Mills et~al.(2021)Mills, Sivarajah, Scholten, and Duncan]{Mills2021application}
Daniel Mills, Seyon Sivarajah, Travis~L. Scholten, and Ross Duncan.
\newblock Application-{M}otivated, {H}olistic {B}enchmarking of a {F}ull {Q}uantum {C}omputing {S}tack.
\newblock \emph{{Quantum}}, 5:\penalty0 415, March 2021.
\newblock ISSN 2521-327X.
\newblock \doi{10.22331/q-2021-03-22-415}.
\newblock URL \url{https://doi.org/10.22331/q-2021-03-22-415}.

\bibitem[Resch and Karpuzcu(2021)]{10.1145/3464420}
Salonik Resch and Ulya~R. Karpuzcu.
\newblock Benchmarking quantum computers and the impact of quantum noise.
\newblock \emph{ACM Comput. Surv.}, 54\penalty0 (7), 07 2021.
\newblock ISSN 0360-0300.
\newblock \doi{10.1145/3464420}.
\newblock URL \url{https://doi.org/10.1145/3464420}.

\bibitem[Lubinski et~al.(2023{\natexlab{a}})Lubinski, Johri, Varosy, Coleman, Zhao, Necaise, Baldwin, Mayer, and Proctor]{Lubinski_etal_2021}
Thomas Lubinski, Sonika Johri, Paul Varosy, Jeremiah Coleman, Luning Zhao, Jason Necaise, Charles~H. Baldwin, Karl Mayer, and Timothy Proctor.
\newblock Application-oriented performance benchmarks for quantum computing, 2023{\natexlab{a}}.

\bibitem[Lubinski et~al.(2024)Lubinski, Goings, Mayer, Johri, Reddy, Mehta, Bhatia, Rappaport, Mills, Baldwin, Zhao, Barbosa, Maity, and Mundada]{lubinski2024quantum}
Thomas Lubinski, Joshua~J. Goings, Karl Mayer, Sonika Johri, Nithin Reddy, Aman Mehta, Niranjan Bhatia, Sonny Rappaport, Daniel Mills, Charles~H. Baldwin, Luning Zhao, Aaron Barbosa, Smarak Maity, and Pranav~S. Mundada.
\newblock Quantum algorithm exploration using application-oriented performance benchmarks, 2024.

\bibitem[Fin\v{z}gar et~al.(2022)Fin\v{z}gar, Ross, Holscher, Klepsch, and Luckow]{Finzgar_2022}
Jernej~Rudi Fin\v{z}gar, Philipp Ross, Leonhard Holscher, Johannes Klepsch, and Andre Luckow.
\newblock {QUARK}: A framework for quantum computing application benchmarking.
\newblock In \emph{2022 {IEEE} International Conference on Quantum Computing and Engineering ({QCE})}. {IEEE}, 9 2022.
\newblock \doi{10.1109/qce53715.2022.00042}.
\newblock URL \url{https://doi.org/10.1109\%2Fqce53715.2022.00042}.

\bibitem[Bowles et~al.(2024)Bowles, Ahmed, and Schuld]{bowles2024better}
Joseph Bowles, Shahnawaz Ahmed, and Maria Schuld.
\newblock Better than classical? the subtle art of benchmarking quantum machine learning models, 2024.

\bibitem[Ladd et~al.(2010)Ladd, Jelezko, Laflamme, Nakamura, Monroe, and O'Brien]{Ladd2010}
T.~D. Ladd, F.~Jelezko, R.~Laflamme, Y.~Nakamura, C.~Monroe, and J.~L. O'Brien.
\newblock Quantum computers.
\newblock \emph{Nature}, 464\penalty0 (7285):\penalty0 45--53, 03 2010.
\newblock ISSN 1476-4687.
\newblock \doi{10.1038/nature08812}.
\newblock URL \url{https://doi.org/10.1038/nature08812}.

\bibitem[Kiwit et~al.(2023)Kiwit, Marso, Ross, Riofrío, Klepsch, and Luckow]{kiwit2023applicationoriented}
Florian~J. Kiwit, Marwa Marso, Philipp Ross, Carlos~A. Riofrío, Johannes Klepsch, and Andre Luckow.
\newblock Application-oriented benchmarking of quantum generative learning using quark, 2023.

\bibitem[Benedetti et~al.(2019{\natexlab{a}})Benedetti, Lloyd, Sack, and Fiorentini]{Benedetti_2019_2}
Marcello Benedetti, Erika Lloyd, Stefan Sack, and Mattia Fiorentini.
\newblock Parameterized quantum circuits as machine learning models.
\newblock \emph{Quantum Science and Technology}, 4\penalty0 (4):\penalty0 043001, 11 2019{\natexlab{a}}.
\newblock \doi{10.1088/2058-9565/ab4eb5}.
\newblock URL \url{https://doi.org/10.1088/2058-9565/ab4eb5}.

\bibitem[Schuld and Petruccione(2021)]{schuld_machine_2021}
Maria Schuld and Francesco Petruccione.
\newblock \emph{Machine {Learning} with {Quantum} {Computers}}.
\newblock Quantum {Science} and {Technology}. Springer International Publishing, Cham, 2021.
\newblock \doi{10.1007/978-3-030-83098-4}.
\newblock URL \url{https://link.springer.com/10.1007/978-3-030-83098-4}.

\bibitem[Benedetti et~al.(2019{\natexlab{b}})Benedetti, Garcia-Pintos, Perdomo, Leyton-Ortega, Nam, and Perdomo-Ortiz]{benedetti2019generative}
Marcello Benedetti, Delfina Garcia-Pintos, Oscar Perdomo, Vicente Leyton-Ortega, Yunseong Nam, and Alejandro Perdomo-Ortiz.
\newblock A generative modeling approach for benchmarking and training shallow quantum circuits.
\newblock \emph{npj Quantum Information}, 5\penalty0 (1):\penalty0 1--9, 2019{\natexlab{b}}.

\bibitem[Lloyd and Weedbrook(2018)]{PhysRevLett.121.040502}
Seth Lloyd and Christian Weedbrook.
\newblock Quantum generative adversarial learning.
\newblock \emph{Phys. Rev. Lett.}, 121:\penalty0 040502, 07 2018.
\newblock \doi{10.1103/PhysRevLett.121.040502}.
\newblock URL \url{https://link.aps.org/doi/10.1103/PhysRevLett.121.040502}.

\bibitem[Riofrío et~al.(2023)Riofrío, Mitevski, Jones, Krellner, Vučković, Doetsch, Klepsch, Ehmer, and Luckow]{riofrio2023performance}
Carlos~A. Riofrío, Oliver Mitevski, Caitlin Jones, Florian Krellner, Aleksandar Vučković, Joseph Doetsch, Johannes Klepsch, Thomas Ehmer, and Andre Luckow.
\newblock A performance characterization of quantum generative models, 2023.

\bibitem[Hansen et~al.(2019)Hansen, Akimoto, and Baudis]{hansen2019pycma}
Nikolaus Hansen, Youhei Akimoto, and Petr Baudis.
\newblock {CMA-ES/pycma} on {G}ithub.
\newblock Zenodo, DOI:10.5281/zenodo.2559634, February 2019.
\newblock URL \url{https://doi.org/10.5281/zenodo.2559634}.

\bibitem[Goodfellow et~al.(2014)Goodfellow, Pouget-Abadie, Mirza, Xu, Warde-Farley, Ozair, Courville, and Bengio]{NIPS2014_5ca3e9b1}
Ian Goodfellow, Jean Pouget-Abadie, Mehdi Mirza, Bing Xu, David Warde-Farley, Sherjil Ozair, Aaron Courville, and Yoshua Bengio.
\newblock Generative adversarial nets.
\newblock In Z.~Ghahramani, M.~Welling, C.~Cortes, N.~Lawrence, and K.Q. Weinberger, editors, \emph{Advances in Neural Information Processing Systems}, volume~27. Curran Associates, Inc., 2014.
\newblock URL \url{https://proceedings.neurips.cc/paper/2014/file/5ca3e9b122f61f8f06494c97b1afccf3-Paper.pdf}.

\bibitem[Schuld et~al.(2019)Schuld, Bergholm, Gogolin, Izaac, and Killoran]{PhysRevA.99.032331}
Maria Schuld, Ville Bergholm, Christian Gogolin, Josh Izaac, and Nathan Killoran.
\newblock Evaluating analytic gradients on quantum hardware.
\newblock \emph{Phys. Rev. A}, 99:\penalty0 032331, 03 2019.
\newblock \doi{10.1103/PhysRevA.99.032331}.
\newblock URL \url{https://link.aps.org/doi/10.1103/PhysRevA.99.032331}.

\bibitem[Kingma and Ba(2017)]{kingma2017adam}
Diederik~P. Kingma and Jimmy Ba.
\newblock Adam: A method for stochastic optimization, 2017.

\bibitem[Jain(1991)]{jain1991art}
R.~Jain.
\newblock \emph{{The art of computer systems performance analysis: techniques for experimental design, measurement, simulation, and modeling}}.
\newblock Wiley New York, 1991.

\bibitem[Amico et~al.(2023)Amico, Zhang, Jurcevic, Bishop, Nation, Wack, and McKay]{amico2023defining}
Mirko Amico, Helena Zhang, Petar Jurcevic, Lev~S. Bishop, Paul Nation, Andrew Wack, and David~C. McKay.
\newblock Defining standard strategies for quantum benchmarks, 2023.

\bibitem[Wang et~al.(2022)Wang, Guo, and Shan]{wang2022sok}
Junchao Wang, Guoping Guo, and Zheng Shan.
\newblock Sok: Benchmarking the performance of a quantum computer.
\newblock \emph{Entropy}, 24\penalty0 (10):\penalty0 1467, 2022.

\bibitem[Eisert et~al.(2020)Eisert, Hangleiter, Walk, Roth, Markham, Parekh, Chabaud, and Kashefi]{Eisert2020}
Jens Eisert, Dominik Hangleiter, Nathan Walk, Ingo Roth, Damian Markham, Rhea Parekh, Ulysse Chabaud, and Elham Kashefi.
\newblock Quantum certification and benchmarking.
\newblock \emph{Nature Reviews Physics}, 2\penalty0 (7):\penalty0 382--390, 07 2020.
\newblock ISSN 2522-5820.
\newblock \doi{10.1038/s42254-020-0186-4}.
\newblock URL \url{https://doi.org/10.1038/s42254-020-0186-4}.

\bibitem[Cross et~al.(2019)Cross, Bishop, Sheldon, Nation, and Gambetta]{Cross2019}
Andrew~W. Cross, Lev~S. Bishop, Sarah Sheldon, Paul~D. Nation, and Jay~M. Gambetta.
\newblock Validating quantum computers using randomized model circuits.
\newblock \emph{Phys. Rev. A}, 100:\penalty0 032328, 09 2019.
\newblock \doi{10.1103/PhysRevA.100.032328}.
\newblock URL \url{https://link.aps.org/doi/10.1103/PhysRevA.100.032328}.

\bibitem[Wack et~al.(2021)Wack, Paik, Javadi-Abhari, Jurcevic, Faro, Gambetta, and Johnson]{wack2021quality}
Andrew Wack, Hanhee Paik, Ali Javadi-Abhari, Petar Jurcevic, Ismael Faro, Jay~M. Gambetta, and Blake~R. Johnson.
\newblock Quality, speed, and scale: three key attributes to measure the performance of near-term quantum computers, 2021.

\bibitem[Koen~Mesman(2022)]{koen_qaoa_benchmarking_2022}
Matthias~Möller Koen~Mesman, Zaid Al-Ars.
\newblock Qpack: Quantum approximate optimization algorithms as universal benchmark for quantum computers.
\newblock \emph{arXiv}, 2022.
\newblock \doi{10.48550/arXiv.2103.17193}.
\newblock URL \url{https://doi.org/10.48550/arXiv.2103.17193}.

\bibitem[Lubinski et~al.(2023{\natexlab{b}})Lubinski, Coffrin, McGeoch, Sathe, Apanavicius, and Neira]{Lubinski_etal_2023}
Thomas Lubinski, Carleton Coffrin, Catherine McGeoch, Pratik Sathe, Joshua Apanavicius, and David E.~Bernal Neira.
\newblock Optimization applications as quantum performance benchmarks, 2023{\natexlab{b}}.

\bibitem[Martiel et~al.(2021)Martiel, Ayral, and Allouche]{Martiel_2021}
Simon Martiel, Thomas Ayral, and Cyril Allouche.
\newblock Benchmarking quantum coprocessors in an application-centric, hard\-ware-agnostic, and scalable way.
\newblock \emph{IEEE Transactions on Quantum Engineering}, 2:\penalty0 1–11, 2021.
\newblock ISSN 2689-1808.
\newblock \doi{10.1109/tqe.2021.3090207}.
\newblock URL \url{http://dx.doi.org/10.1109/TQE.2021.3090207}.

\bibitem[GitHub()]{quarkGithub}
GitHub.
\newblock Quark: A framework for quantum computing application benchmarking.
\newblock \url{https://github.com/QUARK-framework/QUARK}, 2023.

\bibitem[{Qiskit contributors}(2023)]{Qiskit}
{Qiskit contributors}.
\newblock Qiskit: An open-source framework for quantum computing, 2023.

\bibitem[Bergholm et~al.(2018)Bergholm, Izaac, Schuld, Gogolin, Alam, Ahmed, Arrazola, Blank, Delgado, Jahangiri, et~al.]{bergholm2018pennylane}
Ville Bergholm, Josh Izaac, Maria Schuld, Christian Gogolin, M~Sohaib Alam, Shahnawaz Ahmed, Juan~Miguel Arrazola, Carsten Blank, Alain Delgado, Soran Jahangiri, et~al.
\newblock Pennylane: Automatic differentiation of hybrid quantum-classical computations.
\newblock \emph{arXiv preprint arXiv:1811.04968}, 2018.

\bibitem[Zhu et~al.(2022)Zhu, Johri, Bacon, Esencan, Kim, Muir, Murgai, Nguyen, Pisenti, Schouela, Sosnova, and Wright]{PhysRevResearch.4.043092}
Elton~Yechao Zhu, Sonika Johri, Dave Bacon, Mert Esencan, Jungsang Kim, Mark Muir, Nikhil Murgai, Jason Nguyen, Neal Pisenti, Adam Schouela, Ksenia Sosnova, and Ken Wright.
\newblock Generative quantum learning of joint probability distribution functions.
\newblock \emph{Phys. Rev. Res.}, 4:\penalty0 043092, 11 2022.
\newblock \doi{10.1103/PhysRevResearch.4.043092}.
\newblock URL \url{https://link.aps.org/doi/10.1103/PhysRevResearch.4.043092}.

\bibitem[Winston and Moreda(2018)]{QiskitBackends2023}
Erick Winston and Diego Moreda.
\newblock Qiskit backends: What they are and how to work with them.
\newblock \emph{Medium}, 2018.
\newblock URL \url{https://medium.com/qiskit/qiskit-backends-what-they-are-and-how-to-work-with-them-fb66b3bd0463}.

\bibitem[Liu and Wang(2018)]{liu2018}
Jin-Guo Liu and Lei Wang.
\newblock Differentiable learning of quantum circuit born machines.
\newblock \emph{Phys. Rev. A}, 98, 12 2018.
\newblock \doi{10.1103/PhysRevA.98.062324}.
\newblock URL \url{https://-link.aps.org/doi/10.1103/PhysRevA.98.062324}.

\bibitem[Steffen et~al.(2024)Steffen, Chow, Sheldon, and McClure]{IBMerrorrates}
Matthias Steffen, Jerry Chow, Sarah Sheldon, and Doug McClure.
\newblock {IBM Research A} new eagle in the poughkeepsie quantum datacenter: Ibm quantum’s most performant system yet.
\newblock \url{https://research.ibm.com/blog/eagle-quantum-error-mitigation}, 2024.

\bibitem[{IonQ Collaborators}(2024)]{IonQerrorrates}
{IonQ Collaborators}.
\newblock {IonQ Harmony}.
\newblock \url{https://ionq.com/quantum-systems/harmony}, 2024.

\bibitem[Preskill(2018)]{preskill2018}
John Preskill.
\newblock Quantum computing in the nisq era and beyond.
\newblock \emph{Quantum}, 2:\penalty0 79, 2018.
\newblock \doi{10.22331/q-2018-08-06-79}.
\newblock URL \url{https://doi.org/10.22331/q-2018-08-06-79}.

\bibitem[Nielsen and Chuang(2010)]{nielsen_chuang_2010}
Michael~A. Nielsen and Isaac~L. Chuang.
\newblock \emph{Quantum Computation and Quantum Information: 10th Anniversary Edition}.
\newblock Cambridge University Press, 2010.
\newblock \doi{10.1017/CBO9780511976667}.

\bibitem[Oliv et~al.(2022)Oliv, Matic, Messerer, and Lorenz]{oliv2022evaluating}
Marita Oliv, Andrea Matic, Thomas Messerer, and Jeanette~Miriam Lorenz.
\newblock Evaluating the impact of noise on the performance of the variational quantum eigensolver, 2022.

\end{thebibliography}

\end{document}